\documentclass{acmart}
\usepackage[ruled,noend]{algorithm2e}
\newtheorem{asu}{Assumption}
\usepackage{enumitem}
\usepackage{caption}
\usepackage{subcaption}
\usepackage{multirow,multicol}

\AtBeginDocument{%
  \providecommand\BibTeX{{%
    \normalfont B\kern-0.5em{\scshape i\kern-0.25em b}\kern-0.8em\TeX}}}

\copyrightyear{2022}
\acmYear{2022}
\setcopyright{acmlicensed}\acmConference[FAccT '22]{2022 ACM Conference on Fairness, Accountability, and Transparency}{June 21--24, 2022}{Seoul, Republic of Korea}
\acmBooktitle{2022 ACM Conference on Fairness, Accountability, and Transparency (FAccT '22), June 21--24, 2022, Seoul, Republic of Korea}
\acmPrice{15.00}
\acmDOI{10.1145/3531146.3533165}
\acmISBN{978-1-4503-9352-2/22/06}

\begin{document}

\author{Yuhao Du}
\email{yuhaodu@buffalo.edu}
\orcid{0000-0002-2474-8529}
\affiliation{%
  \institution{University at Buffalo}
  \city{Buffalo}
  \country{USA}
  }

\author{Stefania Ionescu}
\orcid{0000-0002-6612-5856}
\email{ionescu@ifi.uzh.ch}
\affiliation{%
  \institution{University of Zurich}
  \city{Zurich}
  \country{Switzerland}}

\author{Melanie Sage}
\orcid{0000-0002-3385-9650}
\email{msage@buffalo.edu}
\affiliation{%
  \institution{University at Buffalo}
  \city{Buffalo}
  \country{USA}}
\email{msage@buffalo.edu}

\author{Kenneth Joseph}
\orcid{0000-0003-2233-3976}
\email{kjoseph@buffalo.edu}
\affiliation{%
  \institution{University at Buffalo}
  \city{Buffalo}
  \country{USA}}

\title{A Data-Driven Simulation of the New York State Foster Care System}

\begin{abstract}
We introduce an analytic pipeline to model and simulate youth trajectories through the New York state foster care system. Our goal in doing so is to forecast how proposed interventions may impact the foster care system's ability to achieve it's stated goals \emph{before these interventions are actually implemented and impact the lives of thousands of youth}. Here, we focus on two specific stated goals of the system: racial equity, and, as codified most recently by the 2018 Family First Prevention Services Act (FFPSA), a focus on keeping all youth out of foster care.  We also focus on one specific potential intervention--- a predictive model, proposed in prior work and implemented elsewhere in the U.S., which aims to determine whether or not a youth is in need of care. We use our method to explore how the implementation of this predictive model in New York would impact racial equity and the number of youth in care.  While our findings, as in any simulation model, ultimately rely on modeling assumptions, we find evidence that the model would not necessarily achieve either goal. Primarily, then, we aim to further promote the use of data-driven simulation to help understand the ramifications of algorithmic interventions in public systems.
\end{abstract}



\maketitle

\section{Introduction}

Youth admitted into foster care in the United States are likely to experience a series of investigations and evaluations. This constant surveillance, often combined with frequent and abrupt shifts in living situations, leave a number of lasting socio-emotional scars \cite{Unrau_multiplePlacement,kolko2010posttraumatic}.  These scars are, moreover, not distributed equally. In particular, significant racial biases exist regarding who enters into foster care \cite{boyd2014african,harris2008decision}.

These and other issues with the American child welfare system have led to a growing movement to abolish it \cite{dettlaff2020not}.  Suggested alternatives from these advocates include reallocating funding to  community-based, localized initiatives not run by government actors \cite{dettlaff2020not}.  Others have, in a more traditional vein for the field of Social Work, instead argued for changing, rather than eradicating the system. However, both abolitionists and reformers generally agree on one point: youth are better off with their own families than in the foster care system  \cite{childrensbureauIssueBriefInHome2014}.

The most important recent development aimed at keeping youth out of care is the \emph{Family First Prevention Services Act (FFPSA)} \cite{FamilyFirst}. The FFPSA, signed into federal law in 2018, overhauled funding in the American child welfare system in order to providing financial incentives to states to keep youth out of foster care, or more formally, to 1) reduce unnecessary separation, and 2) provide family-based service for removed youth in order to encourage family reunification. These incentives include the construction of new benchmarks for numbers of youth in care, and additional funding for family-based services aimed specifically at helping the families of youth who are in care to get to a point where the youth can be returned to them. 

Efforts to effectively use this funding from the FFPSA to reduce the number of youth in foster care face three formidable challenges.  First, the American child welfare system is heavily decentralized, with drastically different policy environments across and often even within states \cite{weigensbergscan}. Changes that are effective at moving towards the goals of the FFPSA in one county or state may therefore be difficult (or even illegal) to enact in others. Second, even in a single jurisdiction, the process of placing a youth into the foster care system is complex. When a potentially maltreated youth is reported to Child Protective Service (CPS), a series of decisions made by case workers and judges are carried out to decide the best path for the youth.  The process is cumbersome and riddled with the potential of both personal and systemic biases \cite{reddenDatafiedChildWelfare2020a,yelickEffectsFamilyStructure2020}. Efforts to discharge youth from foster care can face similar challenges. Finally, as with most social policy settings \cite{rittel1973dilemmas}, attempts to make progress on one goal can often have unexpected negative impacts on others. As we highlight here, for example, interventions aimed making the system more racially equitable may end up leading to substantially more youth in foster care, and not with their families.

Help is needed to tackle these challenges. \citeauthor{ahman2021calculating} \cite{abdurahman2021calculating} argues that implicit in the FFPSA is the assumption that such help will come from the expansion of data collection the bill enables, which will in turn lead to the construction of (predictive) analytical tools that can serve to ameliorate potential problems. Indeed, the use of such tools is already widespread in child welfare, and is only expected to grow with the FFPSA \cite{SamantACLU}.

The present work takes a different bend on how computation can serve a role in social change \cite{abebe2020roles}, arguing that if used carefully, computational models can help us to reason about the best path forward under myriad possible social and policy environments, and various potential interventions. That is, rather than make predictions about youth, we can use computation to help us \emph{understand} the foster care system as a whole, and to \emph{rethink} the potential interventions needed to address new policy goals like those set forth in the FFPSA. 

The present work conducts such an analysis using a two-stage pipeline. We first take a \textit{forensic social science} \cite{mcfarlandSociologyEraBig2015} approach to address the following research question: \textbf{how do existing practices within the New York State child welfare system fare in light of the new stated goals introduced by FFPSA and the existing stated goal of racial equity?} While our methodology generalizes to other contexts, we focus on New York because of our policy and practical expertise in the state. Forensic social science is a methodology in which computational analyses of observational data are conducted in ways that inform and are informed by relevant social theory. Here, using complete data from the child welfare system in New York from 2000-2017 ~\cite{us2015adoption}, we conducted a computational analysis on 1) entry rates of youth into the foster care system, 2) patterns in how long youth stay in care, and 3) the rates at which youth are discharged. We also consider how these quantities differ for white versus Black youth.

Our forensic social science analysis informs the second, and focal, part of our computational pipeline, in which we construct  a \textit{data-driven system dynamics simulation} \cite{gilbertSimulationSocialScientist2005,Sterman2002SystemDM} to analyze a hypothetical intervention into New York's foster care system. Specifically, we analyze how the introduction of an automated risk assessment tool currently in practice in other parts of the country \cite{pmlr-v81-chouldechova18a} would, if implemented in New York, help to address the goals of the FFPSA while maintaining the existing goal of improving racial equity. System dynamics models help formalize how particular entities \emph{flow} through a system over time, conditioned on assumptions about the probability of flowing from one point to another. Here, we model the flow of American youth into and out of foster care. In a data-driven system dynamics model, some of these patterns of flow, and associated probabilities, can be informed by data. Those that cannot be informed by available data can then be set based on assumptions and/or theory. These assumptions, in turn, can then be varied to address certain research questions or to perform robustness checks. 

Our work makes three primary contributions:
\begin{enumerate}
\item We provide quantitative evidence, supporting earlier work, that show that youth stays in foster care in New York can be empirically separated into two classes, long-term and short-term. We find further that the duration of long-term, but not short-term, stays varies significantly across racial lines for all ages. 

\item Informed by these analyses, we develop a data-driven simulation model of the U.S. foster care system, and parameterize it for the study of New York. As a sign of the model's validity, we show that the model can reliably forecast patterns in the number of youth admitted into New York state's foster care system in 2018, given only data from previous years.

\item We use our model to study the impacts of a potential algorithmic intervention in New York. Our observations resonate with the concerned pointed out by \citet{SamantACLU}. Specifically, we find that it is difficult to balance the goals of racial equity and a reduction of youth in care, and that proposed algorithmic interventions---encouraged by the FFPSA in order to achieve it's goals \cite{abdurahman2021calculating}---are not necessarily capable of doing so.
\end{enumerate}

As with any simulation model, our findings rely on modeling assumptions which may be reasonably disagreed with. To facilitate such discussions, we have made our model publicly available.\footnote{https://github.com/yuhaodu/system\_dynamic\_simulation\_FC} However, as we believe our modeling assumptions to be at least within the realm of possibilities, our results suggest that shifting policy landscapes impacts both the validity and utility of using historical administrative dataset to build machine learning models in public sectors. Instead, we argue that machine learning, and computation writ large, may be better served as a tool to facilitate social theory and social policy, rather than to act as explicitly as a decision-making tool, where it is often inserted into the problematic decision loops which expose risks of amplifying existing problems.

\section{Related Work}
Computational social scientists have develop a number of ways to formalize and analyze complex sociotechnical systems \cite{schellingDynamicModelsSegregation1971,gilbertSimulationSocialScientist2005,lazerComputationalSocialScience2009}.  Careful formalization and analysis can serve to illuminate and identify paths towards addressing societal issues \cite{epsteinWhyModel2008}. The benefits and drawbacks of computational modeling of (and for) social systems are perhaps best summarized by \citet{abebe2020roles}, who note that computing can act as a synecdoche, allowing us to think about problems in new ways, and as a rebuttal,  ``clarify[ing] the limits of technical interventions'' \cite[][pg. 256]{abebe2020roles}. Our goal in the present work, similar in some respects to the arguments made by \citet{green2021flaws} but with distinct methods, is to use computation as both a synecdoche and as a rebuttal for the blind reapplication of machine learning methods from one context to another.

Here, we define computational analysis broadly, to include both forensic social science and simulation.  With respect to the former, forensic social science entails the combined use of machine learning and social theory to advance our understanding of social phenomenon, where (in the forensic social science approach) machine learning/advanced statistical analyses are applied an ``atheoretical [approach to] agnostic search for potential explanations,'' and social theory is ``a focusing device that identifies which constructs are to be selected and formed from the millions of possible [analyses]'' \cite{mcfarlandSociologyEraBig2015}, (pg. 10).  We briefly touch on related work in the context of quantitative studies of American Foster Care System that inform our forensic social science analysis, and the simulations of sociotechnical systems that compliment and/or inform the work presented here.

\subsection{Quantitative Study of the American Foster Care System}

Quantitative studies of foster care are routinely conducted by and for policy makers, case workers and researchers. Given the rich administrative datasets curated in this context, quantitative studies have been carried out to both understand the different aspects of the system and to provide assistance with decision making. 

With respect to quantitative work that assists decision making, we direct the reader to a number of recent literature reviews \cite{Saxena2020,SamantACLU} around algorithms used within the U.S. Child Welfare System. Most notably here,  \citet{pmlr-v81-chouldechova18a} build and implement a machine learning model called AFST in Allegheny County, Pennsylvania that helps case workers decide whether or not to screen in reported cases of child abuse for further analysis. This academic work is complemented by a number of public reports as well (e.g. \cite{Jeremy_report}).  

A number of other researchers have used quantitative methods to critique existing practices within the U.S. foster care system (e.g. \cite{okpychReceiptIndependentLiving2015}), to better understand associations with service allocation to youth (e.g. \cite{yan2021computational,lee2008comparing}), to analyze racial disparities within the system (e.g. \cite{wulczyn2007racial,Cheng_CHI_2021,keddell2019algorithmic}), and to criticize the use of automatic decision tools in foster care system (e.g. \cite{roberts2019digitizing,De-Arteaga2020,NCCPR}). Our work compliments these efforts, both in its use of quantitative methods to explore youth lengths of stay in new ways, and to use quantitative methods to critique existing practices. 

\subsection{Social Simulation}

The present work uses a simulation methodology particularly well-suited for our work: \emph{data-driven simulation}. The term data-driven simulation encapsulates a broad range of computing techniques which use relevant data to make educated predictions about what might happen in a situation for which complete data cannot be obtained. Data-driven simulation has seen increasing use in the FAccT-aligned community \cite{hu2019disparate,milli2019social,d2020fairness,goldhaber2012evaluating,milli2019social,ionescu2021agent} As noted above, we use a specific form of data-driven simulation, System Dynamics modeling \cite{Sterman2002SystemDM}. Perhaps most relevant to our work, then, \citet{Martin2020ExtendingTM} argues that instead of focusing on mathematical-based interventions on opaque algorithms and/or models, using community based system dynamics modeling to place the algorithm/model into the social context is a better way to understand the long-term impact of algorithms/model. While we aspire to community-based methods, our current work relies only on our existing knowledge of and experiences as practitioners within child welfare. 

Finally, while significantly distinct in focus, it is worth noting that there is other work using simulation in the foster care context. Specifically, \citet{fowler2020scaling} use a system dynamic model to test the impact of scaling up a policy to provide long-term rental subsidies for foster care family, and \citet{goldhaber2012evaluating} use decision-analytic model in support of Child Welfare policymakers considering implementing evidence-based interventions. 


\section{A Brief Overview of How Youth Enter Foster Care in America} \label{sec:usfc}

\begin{figure}[t]
\centering
\includegraphics[width = 1\linewidth]{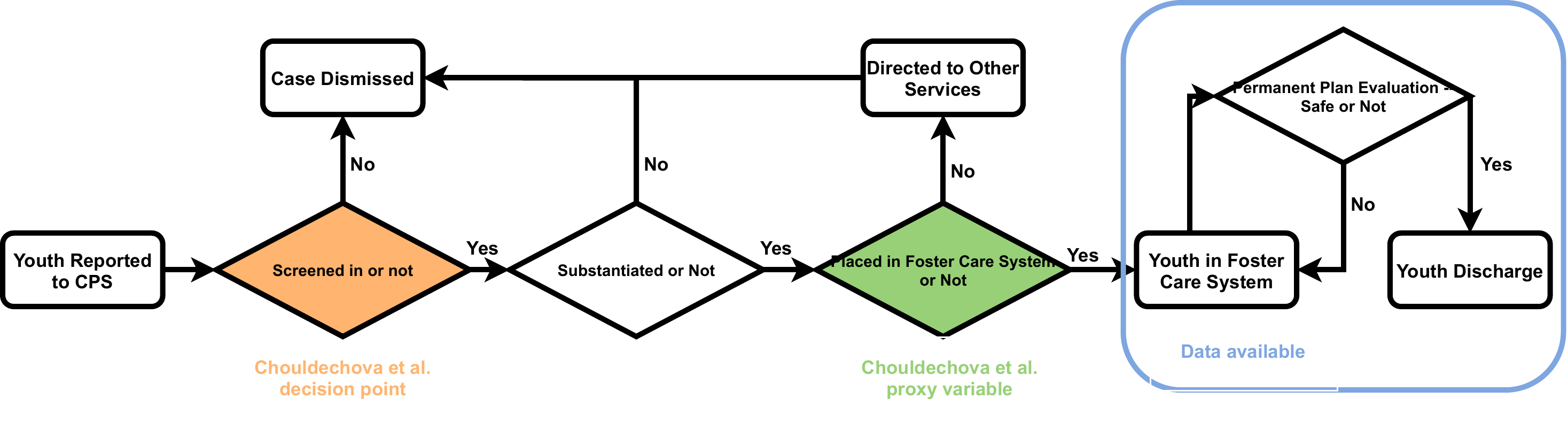}
\caption{The standard decision-making pipeline for a youth reported to Child Protective Services in the United States. Diamonds represent decision points in the system, squares are states that youth may be in at a given time. The blue box represents the portion of the decision pipeline informed by AFCARS data. The orange and green boxes represent, respectively, the decision point for the algorithmic intervention considered in the present work, and the proxy variable used to train the proposed model.}
\label{fig:decision_pipeline}
\end{figure}

Our work assumes a slightly simplified process model of youth trajectories through the U.S. foster care system introduced by \citet{Sterman2002SystemDM}, and visualized in Figure~\ref{fig:decision_pipeline}.
The first step in the process of placing a youth in foster care is a report to Child Protective Services (\emph{CPS}). These reports are handled by a CPS employee who decides whether or not to \emph{screen in} the call, a decision based on myriad factors, e.g. characteristics of the youth, their family, and the local policy environment \cite{pmlr-v81-chouldechova18a}.

If a call is not screened in, the case is dismissed, and the youth exits the system (although notably, their data may not \cite{abdurahman2021calculating}). If a call is screened in, the case is then taken on for further consideration by a CPS case worker. This phase typically includes a more detailed records review of the youth and their family, and a visit to the location relevant to the call (often, the current living situation of the youth). The CPS case worker then decides whether or not the case is \emph{substantiated}, i.e. whether there is evidence that the concern voiced in the original call is true. 

If a case is substantiated, a decision is then made on whether to a) remove the youth from their home, or b) to keep the youth in their home. This decision is typically made by a judge. If the decision is that the youth should be removed from their family, the youth then enters foster care. Once the youth is in foster care, the family is repeatedly re-evaluated for a need to be in the system. More specifically, judges and case workers are expected to return the youth to their family when they are satisfied that it is safe to do so, while also making a backup plan, according to what is referred to as the permanency plan for that youth. When the home is safe again, or alternatively another long-term plan such as guardianship or adoption is available, the youth is then discharged back to a living placement outside of the foster care system.


\section{Data}

Our analysis uses federal administrative data from the {\em Adoption and Foster Care Analysis Reporting System (AFCARS)}~\cite{us2015adoption}, a dataset from the \textit{National Data Archive on Child Abuse and Neglect (NDACAN)}. AFCARS data contains a range of information for all foster youth from all 50 U.S. states, the District of Columbia, and Puerto Rico. This information includes demographic data, like race and gender, and administrative data, like the youth's current placement setting. However, critically, such information is only available for youth who \emph{are in foster care}. Other relevant data --- for example, the number of youth who are reported but not removed, are not in this dataset. As we discuss further below, we can thus only make assumptions about these youth, which we can then vary to emulate different potential real-world settings.

Our analysis uses a particular sample of youth from the full AFCARS dataset. First, as noted above, we focus our analysis only on data from New York state.  Second, our analysis below considers racial disparities in a number of ways. Due to limited data availability leading to imprecise estimates, we focus here only on data for youth who identify (or are identified as) white or Black. This is a significant limitation of the present work that could be alleviated in the future by additional data and more diverse expertise.

\section{Forensic Social Science Analysis}

We conduct a forensic social science analysis to analyze the foster care system in New York with respect to the number of youth in care and racial equity within the system. These analyses are useful both in providing a better understanding of how many youth are in care at any time and for how long they remain in foster care, for seeing how these quantities vary for Black versus white youth, and for informing parameters of our data-drive simulation model.

We conduct an analysis of three quantities: how many (Black vs. white) youth are entering into care (the \emph{entry rate}), how long youth stay in care (\emph{length of stay}), and the rate at which youth are discharged (\emph{discharge rate}). Prior work has analyzed entry rates at the state level, finding potential associations with race and age \cite{russellDemographicsPolicyFoster2015,beckerPredictorsSuccessfulPermanency2007}. Other work analyzing youth in Florida also suggest race is a key factor influencing the length of stay \cite{beckerPredictorsSuccessfulPermanency2007}.  Motivated by this work, we therefore study these quantities split out by youth age in all cases. Moreover, in order to obtain results which are informative for the system dynamics model, we aggregate data at a monthly level.

\subsection{Entry Rate}

To determine entry rates, we extract the total number of admitted youth $n_{r,a}^t$,  where $r$ represents the youth's race (here, Black or white), $a$ represents the age at which they were admitted (rounded down), and $t$ represents the admitted month. For each combination of $r$ and $a$, we calculate the difference of the number of admitted youth between two consecutive months by $\Delta_{r,a}^t = n_{r,a}^{t+1} - n_{r,a}^{t}$. After confirming that this difference is stationary for the vast majority of $(r, a)$ combinations\footnote{Using the Augmented Dickey-Fuller test \cite{dickey1979distribution}, only one p-value was greater than .05, and only two were greater than .001.}, we fit the resulting data to normal distributions $\mathcal{N}^{\Delta_{r,a}}$ for each $\Delta_{r,a}$ along $t$.

\begin{table}[t]
\footnotesize
\centering
\begin{tabular}{|c|c|c|}
\hline
Age & Distribution (Black) & Distribution (white) \\
\hline\hline
1 & $\mathcal{N}(0.25,7.1)$ & $\mathcal{N}(-0.05,6.0)$\\
\hline
5 & $\mathcal{N}(0.15,5.2)$ & $\mathcal{N}(0.01,5.8)$\\
\hline
9 & $\mathcal{N}(0.1,3.8)$ & $\mathcal{N}(0.04,4.1)$\\
\hline
13 & $\mathcal{N}(0.09,5.9)$ & $\mathcal{N}(0.025,4.8)$\\
\hline\hline
\end{tabular}
\caption{Fitted normal distributions for the difference between the number of admitted Black or white youth in consecutive months in the AFCARS dataset}
\label{table:differenceGaussian}
\end{table}

Samples of the estimated distributions for changes in entry rates for white and Black youth at various ages are displayed in Table~\ref{table:differenceGaussian}. Across the visualized quantities, as well as for all ages not displayed, we find no significant differences in the \emph{rate of change} in the number of Black and white youth entering the system. However, consistent with prior work at a national level \cite{wulczyn2007racial}, we observe that Black youth are over-represented relative to white youth within New York.

\subsection{Length of Stay}

\begin{figure}
\centering

\begin{subfigure}[b]{\textwidth}
\centering
\includegraphics[width = 0.8\textwidth]{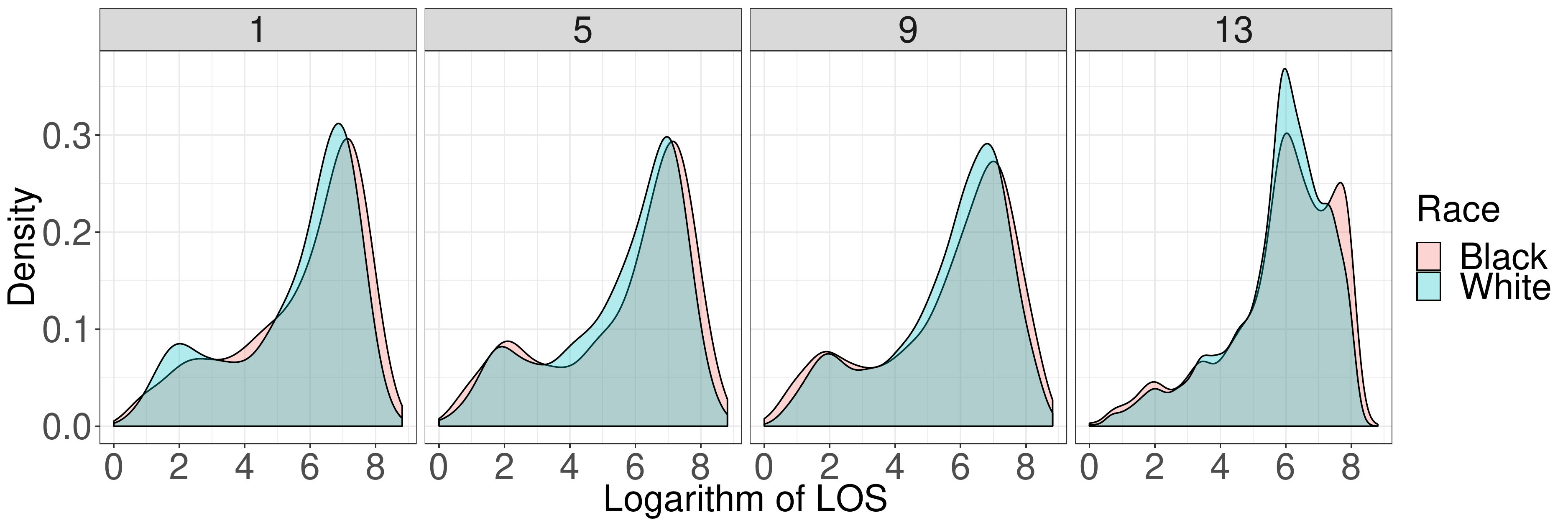}
\caption{
The distributions of log-LOS depending on the youth's race. From left to right, plots correspond to a different age at which the youth was admitted into foster care: 1, 5, 9, and 13. Other ages show similar patterns, and are not displayed}
\label{fig:density}
\end{subfigure}
\begin{subfigure}[b]{\textwidth}
\centering
\includegraphics[width = 0.8\textwidth]{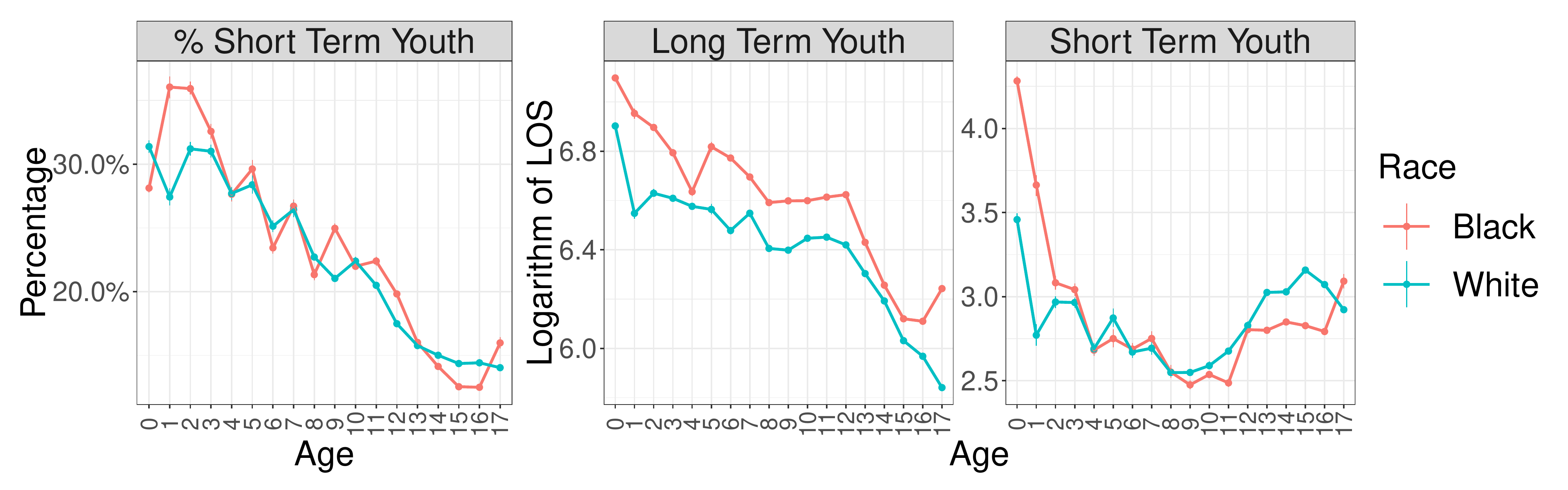}
\caption{On the X-axis is youth's admitted age. The Y-axis of the leftmost plot represents the percentage of short-term youth and the Y-axis at the right two plots represents mean of log-LOS. Cyan lines shows results for white youth, red for Black youth. Data are for youth in New York state in the AFCARS dataset. Confidence interval are 95\% bootstrapped CIs.}
\label{fig:short_long}
\end{subfigure}
\caption{Split of youth according to their LOS}
\label{fig:two_graph}
\end{figure}

Figure~\ref{fig:density} shows the density of the \emph{logarithm of the length of stay (log-LOS)} for white and Black youth admitted at ages 1, 5, 9, and 13. Other ages show similar patterns, and thus we omit the results here. From these plots, we notice that the distribution of log-LOS appears to be readily modeled by a mixture of Gaussian distributions. In other words, there appear to be distinct \emph{classes} of stay lengths for foster youth, that align with different distributions of length of stay.

To test this observation statistically, we fit each individual distribution of log-LOS for all combinations of $a$ and $r$ to five different Gaussian mixture models, with the assumed number of distributions $N$ to be either $N=1,2,3,4,5$. Confirming our intuitions from Figure~\ref{fig:density}, we find that $N=2$ best fitted all densities for all age and race combinations according to the AIC score. 

This Gaussian mixture model separates stays in foster care into two groups- \emph{long-term} and \emph{short-term} stays. Long-term stays have longer log-LOS than short-term stay. For example, among white youth at age 10, the median length of a short-term stay is 10 days, while it is 680 days for a long-term stay. Among Black youth at age 10, the median length of a short-term stay is 9 days, and 804 for a long-term stay.

Breaking AFCARS data down further by long-term youth vs short-term youth is revealing.  First, as shown in the left-most plot in Figure~\ref{fig:short_long}, we find that between 10-30\% of admitted Black and white youth are short-term, with that quantity generally (and predictably  \cite{yan2021computational}) decreasing as youth are older. Second, the middle subplot shows that, consistently across age groups, long-term stays for Black youth are longer than those for white youth. Similarly, the rightmost plot in Figure~\ref{fig:short_long}, shows that that from ages 0-2, short-term stays are also longer for Black youth. There is some evidence that short term stays are slightly longer for older white youth, but this finding is inconsistent at older ages (notably, at age 17).


\subsection{Discharge Rate}

We formalize discharge rate as \emph{the percentage of youth that are in care after $t$ months}. We again learn this function separately for all combinations of $a$ and $r$, as well as for long/short-term foster care youth. We then extracted the empirical inverse cumulative distribution of LOS for long/short-term stays for youth of different demographics. Finally, we learn a non-parametric discharge function using a linear interpolation to fit the empirical inverse cumulative distribution. This \emph{discharge function} $\mathcal{D}_{r,a}^{l}(d_t)$ gives us the probability of $l$-term youth whose race is $r$ and admitted age is $a$ will still be in foster care after $d_t$ months.  

\begin{figure}[t]
\centering
\includegraphics[width =0.5\linewidth]{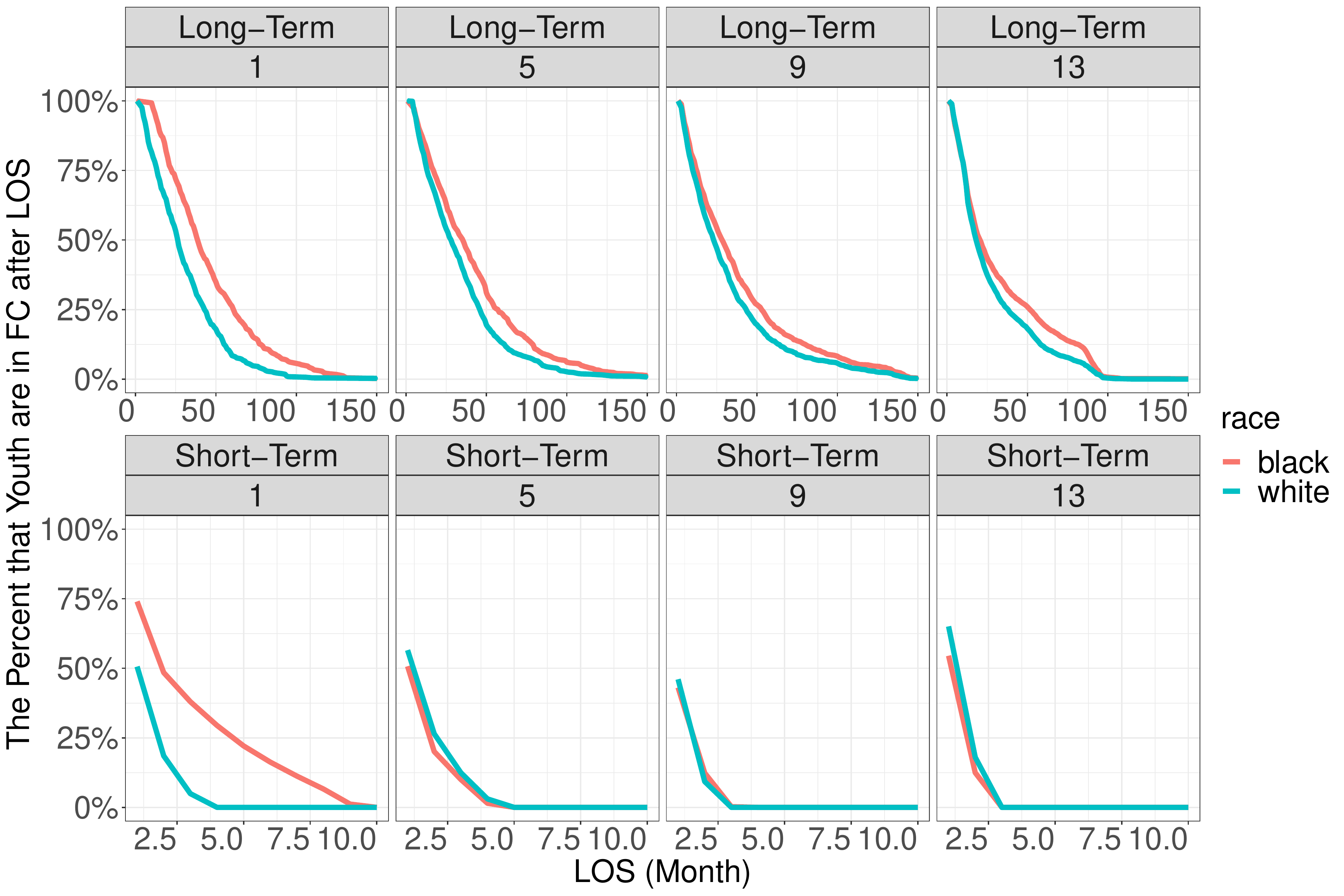}
\caption{On the X-axis is youth length-of-stay. The Y-axis represents the probability that a given youth will still be in foster care after spending the given number of months in foster care. The first row shows the results for long-term youth while second row shows the results for short-term youth. Cyan lines shows results for white youth, red for Black youth. Data are for youth in New York state in the AFCARS dataset}
\label{fig:dischargeRate}
\end{figure}

A representative portion of the results from this analysis are presented in Figure~\ref{fig:dischargeRate}, which shows the expected length of stay for Black and white youth (admitted at ages 1,5,9 and 13) separated out into long-term and short-term  stays for youth. Results mirror findings above, namely, that long-term stays for Black youth, and short-term stays for the youngest Black, are both longer than for the corresponding white youth.

\subsection{Summary of Findings and Linking Back to Relevant Social Theory}

Our empirical analysis reveals that 1) there is a relatively constant rate of entry into the NYS foster care system, 2) that youth can generally be categorized into having a \emph{short-term} or a \emph{long-term} stay in care, and that 3) Black youth who have long-term stays, and the youngest Black youth with short-term stays, remain on average for longer in the foster care system than their white counterparts.

In a forensic social science analysis, it is important to not only guide analyses with prior social science research, but also to tie back to this work once the analysis has been conducted.  For our work, the most critical observation is the bimodal distribution in length of stay across all ages and races that differentiate youth into what we call \emph{long-term} youth \cite{Rosenberg_long_term} and \emph{short-term} youth \cite{sankaran2016easy}.

Rather than social theory, however, it is practical knowledge (derived from our experiences in the system) that help us to contextualize this finding. In particular, we understand short-term stays, which are often the result of brief parental incarceration, or short-term concerns about child safety, to be a distinct class of events relative to longer-term stays, which are often the result of structural factors.

This link between long-term stays and structural factors can, in turn, help us to better understand a more novel finding from our work: differences in length of long-term stays, but not short-term stays, for Black vs. white youth. Both critical race theory in general, and critical veins of social work research, emphasize that race is a socially constructed concept intertwined with structural factors. One such factor in particular is poverty. And, although poverty is not always  indicator for removal, it is often a proxy for reunification. For instance, children are not removed due to homelessness, but housing is often a requirement for a child's return home if they enter foster care \cite{shinn2015poverty}.  Second, single motherhood, also  linked to fewer financial resources, also predicts slower reunification \cite{harris2003interaction}. Third, Black families have disparate access to supportive services, including employment, substance use, and mental health treatment \cite{knott2012african}. Finally, Black youth are less likely to move from foster care into alternative plans, such as adoption and guardianship \cite{akin2011predictors}.  These four mechanisms are examples of how structural factors can serve to create long-term barriers to reunification for Black youth, more so than white youth.


Finally, having contextualized our findings within the literature, we turn to their policy implications. With respect to the goals of the FFPSA, and of improving life for foster youth more broadly, it is useful to distinguish between youth who are more likely to have short and long-term stays in foster care. Short term youth are typically in care for less than a month before being returned to their families. This makes them likely candidates, under the auspice of the FFPSA, to never have been removed at all \cite{van2010predictors,garcia2014three,sankaran2016easy}. Additionally, we note that data useful for policy in child welfare need not be highly coercive or individualized to be useful; rather,  as others have noted \cite{rodriguez2019bridging}, even analyses (or acknowledgement) of aggregate rates of shorter-term versus longer-term foster care stays may appraise agencies to needs related to distribution of services, and help agencies assess the success of their racial equity efforts.


In this sense, algorithms and simulations like those described below may serve as a policy function precisely via the analysis of aggregate rates and measures of allocation. 
To this end, we now turn to how our insights on aggregate measures from this section can be further used to construct a model that simulates the number of youth in foster care in New York state in the federal fiscal year 2018 (FY 2018), under current conditions and in response to an algorithmic intervention.

\section{Data-driven Simulation Analysis}\label{sec:model}

Informed by the discussion and findings above, we proceed to our analysis of a hypothetical algorithmic intervention in the New York state foster care system.  Two aspects in particular from our forensic social science analysis carry over. First, our simulation model explicitly differentiates between youth with short-term versus long-term stays. 
Second, we use a variety of statistics from the work above to inform parameterization of our model. In what follows, we first provide an overview of our model which simulates the trajectory of youth within the New York state foster care system, and show it produces reliable forecasts for the number of youth in foster care.  We then use this model model to investigate the potential impacts of replacing human decision-makers in the screening portion of the child welfare decision pipeline with an algorithm.

\subsection{Model Overview}

The goal of our model is to simulate, for each future month $t$ in $\mathcal{T}_{future}$,  the number of long- and short-term youth (denoted by $l$) having a select combination of race $r$ and age $a$ who are still in foster care system. We denote this number as $N_{r,a}^{l,t}$, and simulate it at at each stage of the decision pipeline outlined in Figure~\ref{fig:decision_pipeline}.

To parameterize the portion of the model where youth are in foster care, we use data from AFCARS for each month until the last month of FY 2017, i.e., each month in $\mathcal{T}_{history}$. Where data is not available to inform the model parameters (that is, when youth are not in care and thus outside the purview of the data in AFCARS), we model a range of plausible assumptions about decision pipelines of foster care system, and assess results across these possible settings. More precisely, we choose our parameters using the following steps:   
\begin{itemize}[nolistsep]
    \item Extract the number of admitted youth $n_{r,a}^{t}$ from AFCARS, where $t \in \mathcal{T}_{history}$.
    \item Extracted $\mathcal{N}^{\Delta_{r,a}}$ and use it to simulate the number of admitted youth $n_{r,a}^{t}$, where $t \in \mathcal{T}_{future}$.
    \item Extract the long-/short-term youth proportion $\mathcal{P}_{r,a}^{l}$. 
    \item Extract the discharge rate $\mathcal{D}_{r,a}^{l}$.
\end{itemize}

Using these parameterizations, we simulate the number of youth in care in New York using Algorithm~\ref{algo:simulation}. For this algorithm, we use the fact that the number of youth in foster care in the next month, $N_{r,a}^{l,t}$, equals the total number of youth who were 1) admitted into foster care during previous months and that 2) remained in the system. In turn, the number of youth who remained in the system is given by the product between the number of admitted youth in the previous month and the discharge rate (line 8-12 in Algorithm~\ref{algo:simulation}). 



Finally, in addition to modeling how many youth are admitted, the proposed intervention we study also requires assumptions about 1) how many youth are \emph{not} admitted (i.e. that are screened out), as well as 2) how many youth cases are screened in. Given the literature discussed above \cite{Rosenberg_long_term,sankaran2016easy}, we assume the following:

\begin{asu} \label{asu:equal}
\vspace{-.2em}
Screen in and substantiation decisions only depend on a youth's race and whether their stay is designated as one who will have a short-term or long-term stay 
\vspace{-.2em}
\end{asu}

Assumption~\ref{asu:equal} says that every reported youth of a given race with a short or long-term stay will have the same probability to be screened in and substantiated. As discussed further below, this assumption notably assumes that substantiation rate does not vary with screen in rates, and vice versa. We fixed this value as introduced in Table~\ref{table:parameter}. Thus, for example, given the simulated number of admitted long-term Black youth at month $t$, $n_{b,a}^{lo,t}$, we can obtain the number of \textit{screened-in} long-term Black youth at month $t$,  $\dot{n}_{b,a}^{lo,t} =  \frac{n_{b,a}^{lo,t}} {R^{sub,lo}_{b}}$, and the number of \textit{reported} long-term Black youth at month $t$, $\ddot{n}_{b,a}^{lo,t} =  \frac{n_{b,a}^{lo,t}}{R^{sub,lo}_{b} \cdot R^{scr,lo}_{b} }$.


\begin{table}
\centering
\footnotesize
\caption{\label{table:parameter}Tabular description of parameters involved in the construction of our simulation model. The red value of the parameter is the default value}
\begin{tabular} {||p{3cm}| p{3cm} | p{3cm}| p{3cm} ||}
\hline
Parameter & Values Taken  & Parameter & Values Taken \\
\hline \hline
 $R^{scr,lo}_{b}$: Screen in rate of \newline long-term Black youth & $70\%$ &
$R^{scr,sh}_{b}$: Screen in rate of short-term Black youth & $6\%$,$8\%$,{\color{red}$10\%$}\\ 
\hline
$R^{scr,lo}_{w}$: Screen in rate of long-term white youth & $70\%$  & 
$R^{scr,sh}_{w}$: Screen in rate of short-term white youth& $10\%$ \\ 
\hline
$R^{sub,lo}_{b}$: Substantiate rate of long-term Black youth  & $70\%$ & 
$R^{sub,sh}_{b}$: Substantiate rate of short-term Black youth  & $10\%$ \\ 
\hline
$R^{sub,lo}_{w}$: Substantiate rate of long-term white youth  & $70\%$  & 
$R^{sub,sh}_{w}$: Substantiate rate of short-term white youth  & $10\%$ \\ 
\hline
$ \mathcal{N}^{lo}(\mu_{lo}, \sigma_{lo})$: Distribution of feature of long-term youth  & $([1,1],{\color{red}[2,2]},[10,10],\newline [[1,0][0,1]])$  & 
$ \mathcal{N}^{sh}(\mu_{sh}, \sigma_{sh})$: Distribution of feature of short-term youth & $([0,0],[[1,0][0,1]])$ \\
\hline
\end{tabular}
\end{table}

\begin{algorithm}[t] 
\begin{multicols}{2}
\footnotesize
\SetAlgoLined
\DontPrintSemicolon
 \KwIn{ $n_{r,a}^{t}$, $\mathcal{P}_{r,a}^{l}$, $\mathcal{D}_{r,a}^{l}$ }
\KwOut{$N_{r,a}^{l,t}$} 
\nl \For{$t \in \mathcal{T}_{future}$}{
\nl \For{ \textrm{l,r,a} \textrm{in} $\mathcal{L} \bigtimes \mathcal{R} \bigtimes{A}$}{
\nl $a^\prime,t^\prime,d_t = a,t,1$ \;
\nl $N_{r,a}^{l,t} = 0$ \;
\nl \While{$a^\prime \geq 0$ and $t^\prime \geq  \textrm{Nov\;2000}$ }{
\nl $\widehat{\Delta}_{r,a}^{l,t} = \mathcal{P}_{r,a^\prime}^{l} \cdot n_{r,a^\prime}^{t^\prime}$ \;
\nl {\color{red}\textit{// \# of admitted long/short-term youth}} \;
\nl $\widehat{\Delta}_{r,a}^{l,t} =\widehat{\Delta}_{r,a}^{l,t} \cdot \mathcal{D}_{r,a}^{l}(d_t) $ \;
{\color{red}\textit{// \# of remaining long/short-term youth after $d_t$ months }} \;
\nl $N_{r,a}^{l,t} \mathrel{+}=\widehat{\Delta}_{r,a}^{l,t}$ \;
\nl $d_t = d_t + 1$ \;
\nl $t^\prime = t^\prime - 1$ \; 
\nl \If{$(d_t - 6) | 12$}{
\nl $a^\prime = a^\prime - 1$}}}}
\caption{Simulation}
\label{algo:simulation}
\columnbreak
\textbf{Notations:} \;
$r:$ Race of youth \;
$a:$ Age of youth \;
$l:$ Long and short-term membership of youth \;
$n_{r,a}^{t}:$ The number of admitted youth at month t \;
$N_{r,a}^{l,t}:$ The number of remaining youth of at month t \;
$\mathcal{P}_{r,a}^{l}:$ Percentage of long-/short-term youth of total \;\quad \quad \quad admitted youth \;
$\mathcal{D}_{r,a}^{l}:$ The discharge rate \;
\end{multicols}
\end{algorithm}



\subsection{Validation of the Base Model}

Figure~\ref{fig:simulation} shows that, given data from 2017 and before, our model generates forecasts for the number of youth in foster care in New York state in 2018 that are in line with real data. The figure shows model predictions (triangle points) for Black (red) and white (cyan) youth separately, comparing simulated estimates aggregated over long-term and short-term youth of all ages.  It compares these predictions to the ground truth values (circle points) in the AFCARS dataset. The Pearson correlation between the number of simulated Black youth and the number of actual Black youth in foster care is $0.98$ ($p  < 0.001$). The Pearson correlation between the number of simulated white youth and the number of real white youth in foster care is also $0.98$ ($p  < 0.001$). And the Pearson correlation between the proportion of Black youth in simulated foster care system and the proportion of Black youth in the actual data is $0.78$ ($p= 0.004$). The results show that our simulation model is a reliable starting point from which to model potential interventions resulting from the implementation of the FFPSA in 2018.
    
\subsection{Extending the Model to Investigate the Effects of an Algorithmic Intervention}\label{sec:algo}

Our hypothetical intervention is based on the work of \citet{pmlr-v81-chouldechova18a}, who develop a model to assist screening decisions in Allegheny County, PA. We stress that \citet{pmlr-v81-chouldechova18a} do not intend for their model to replace humans, and that there are important effects of maintaining a human in the loop during this decision process \cite{Cheng_CHI_2021}. However, it is nonetheless informative to study the simplified case where the model does, in fact, fit this role. Below, we detail our (simulated) implementation of their algorithm, and the additional model parameters we vary to explore how the model responds to different potential context in which it is deployed.

\subsubsection{Simulating Training and Deployment of the Algorithm}

To build the machine learning algorithm to identify the youth who need to be screened in, \citet{pmlr-v81-chouldechova18a} use what we will call the \emph{profiles} of youth. These profiles include demographics, behavioral health records and past history in care (among other variables) as features, and well as whether or not the youth end up in the foster care as the outcome to be predicted. 

To simulate the training procedure for their model, we define (and simplify) the \emph{features} given to the model to be continuous values drawn from a Gaussian distribution. Given our knowledge above that youth with short- and long-term stays are likely to have distinct reasons for being brought into care, we further assume that their profiles are generated from two different distributions. Profiles of long-term youth are generated from $\mathcal{N}^{lo}(\mu_{lo},\sigma_{lo})$ while profiles of short-term youth are generated from  $\mathcal{N}^{sh}(\mu_{sh},\sigma_{sh})$. Parameters of these two distributions are shown in Table~\ref{table:parameter}. 

We define the \emph{label} the model is trained on using the (simulated) decisions at the substantiation phase. Notably, as shown in Figure~\ref{fig:decision_pipeline}, \citeauthor{pmlr-v81-chouldechova18a}'s \citeyear{pmlr-v81-chouldechova18a} model makes predictions at the screen-in stage of the child welfare pipeline. Thus, the label used is a downstream proxy assumed to be less racially biased. In this setup, however, \cite{pmlr-v81-chouldechova18a} therefore assume that if youth end up in foster care at the last decision made in the decision pipeline, then the youth should be screened in at the first decision, and vice versa. 

We simulate the impact of introducing the model from \citet{pmlr-v81-chouldechova18a} as if it was trained using a dataset generated before November 2017, and implemented during the following year. To do so, we construct the training set by using profiles of admitted youth as positive samples and using profiles of reported but not admitted youth as negative samples. After the training dataset is constructed, we use it to train a logistic regression model as a screen-in recommendation tool. \footnote{Training data was balanced by down-sampling negative samples to approximate how machine learning models are trained in real world. Note that we don't modify hyperparameters of the model (e.g. the decision threshold) because this induces yet another implicit value judgement which is out of the scope of our main focus in the current work.}
In our simulation, we model the deployment of this machine learning model as having complete autonomy over decision-making. To be more specific, after deploying the algorithm, step (8) in Algorithm~\ref{algo:simulation} is changed to send the profiles of $\ddot{n}_{r,a}^{l,t}$ reported youth at each month to the deployed algorithms to extract youth screened in. And then, admitted youth are extracted randomly using the substantiate rate introduced in the Table~\ref{table:parameter} following Assumption~\ref{asu:equal}, because substantiation decisions are still made entirely by humans.


\subsubsection{Varying Assumptions to Evaluate the Intervention}

We evaluate the performance of this algorithmic intervention under different assumptions about the underlying foster care system on which it is trained. These variations are summarized in Table~\ref{table:parameter} and detailed below.


\textit{Separability of Short- vs. Long-term Youth Profiles.} We model the difficulty of distinguishing between long-term foster care youth and short-term foster care youth from their profiles as sampling long-term foster care youth's profile from different Gaussian distributions. For example, when long-term foster care youth's profiles are sampled from $\mathcal{N}\sim([10,10],[[1,0],[0,1]])$ and short-term foster care youth's profiles are sampled from $\mathcal{N}\sim([0,0],[[1,0],[0,1]])$, it is much easier for an algorithm to differentiate them,  compared to the situation where long-term foster care youth's profiles are sampled from $\mathcal{N}\sim([1,1],[[1,0],[0,1]])$ and short-term foster care youth's profiles are sampled from $\mathcal{N}\sim([0,0],[[1,0],[0,1]])$. We alter the mean of the feature distribution of long-term foster care youth in ($[10,10],[2,2],[1,1]$) to represent \emph{high} ,\emph{moderate} and \emph{low} separability between youth who will have long versus short term stays, leaving other parameters at their defaults. 

\textit{Modeling Racial Biases in Reporting.}  We model reporting bias across racial lines in the current foster care system by varying the screen-in rates of short-term Black youth. Controlling for the number of admitted Black youth with short-term stays, the lower the screen-in rate, $R_{b}^{scr,sh}$,  for Black youth with short-term stays, the larger the number of Black youth reported. We vary the screen in rate for short-term Black youth in ($6\%$,$8\%$,$10\%$) to represent \emph{high reporting bias}, \emph{low reporting bias}, and \emph{no reporting bias}, leaving other parameters at their defaults.

\subsection{Results of Assessing an Algorithmic Intervention}

\begin{figure}[t]
\centering

\begin{minipage}{0.45\textwidth}
\includegraphics[width = 1\linewidth ,height=5cm ]{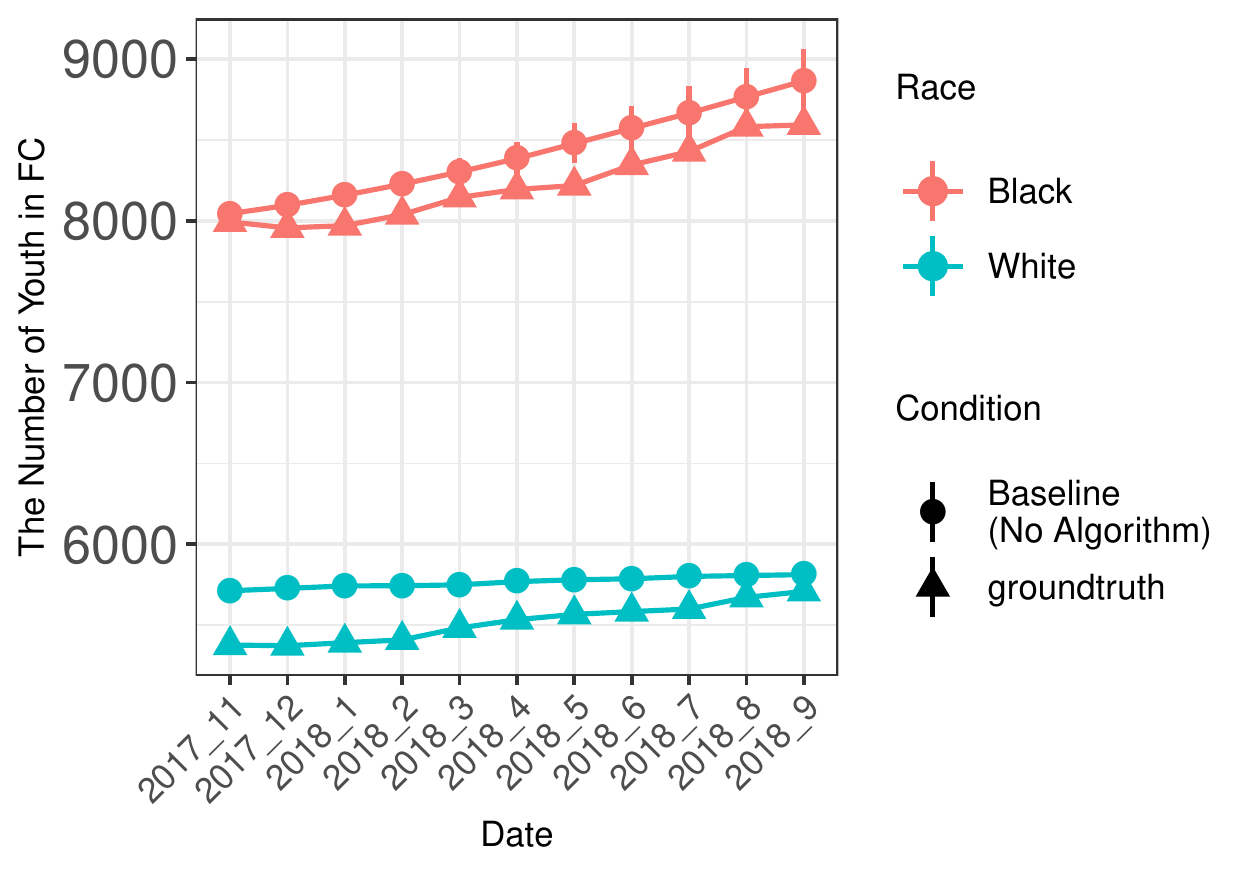}
\caption{
The number (y-axis) of Black (red line) and white (turqouise line) youth in foster care, by month (x-axis), for both the simulated (circles) and real (triangle) foster care data in FY 2018. Real data is derived from the AFCARS dataset for New York state, simulated data from our general simulation model.}
\label{fig:simulation}
\end{minipage}
\hspace{0.2cm}
\begin{minipage}{0.45\textwidth}
\centering
\includegraphics[width = 1\linewidth,height=5cm]{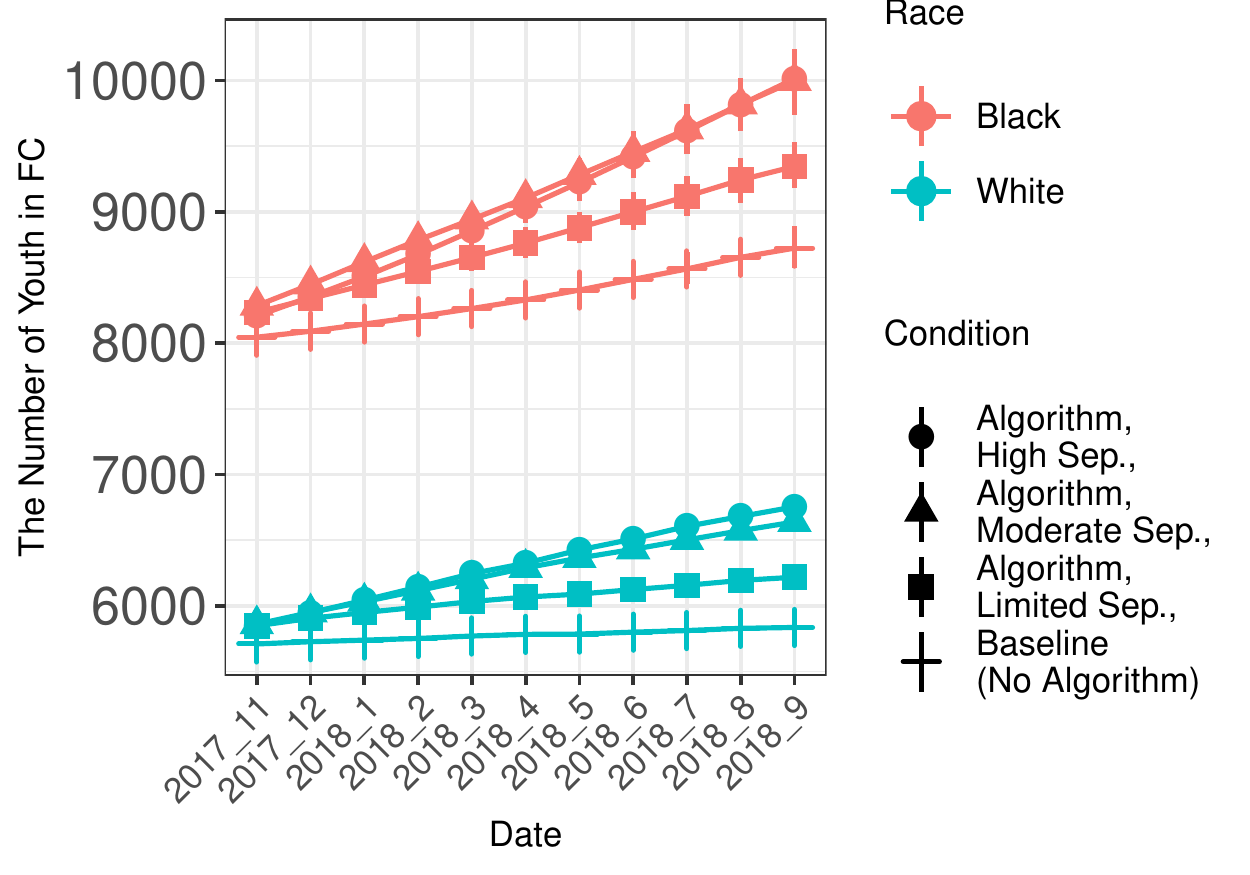}
\caption{ The simulated number of Black and white youth in foster care by month for different screen-in procedures: without an algorithm and with an algorithm but different levels of separability between short-term and long-term youth profiles.}
\label{fig:difficulty}
\end{minipage}
\end{figure}

\begin{figure}[t]
\centering
\begin{subfigure}[b]{\textwidth}
\includegraphics[width=\linewidth]{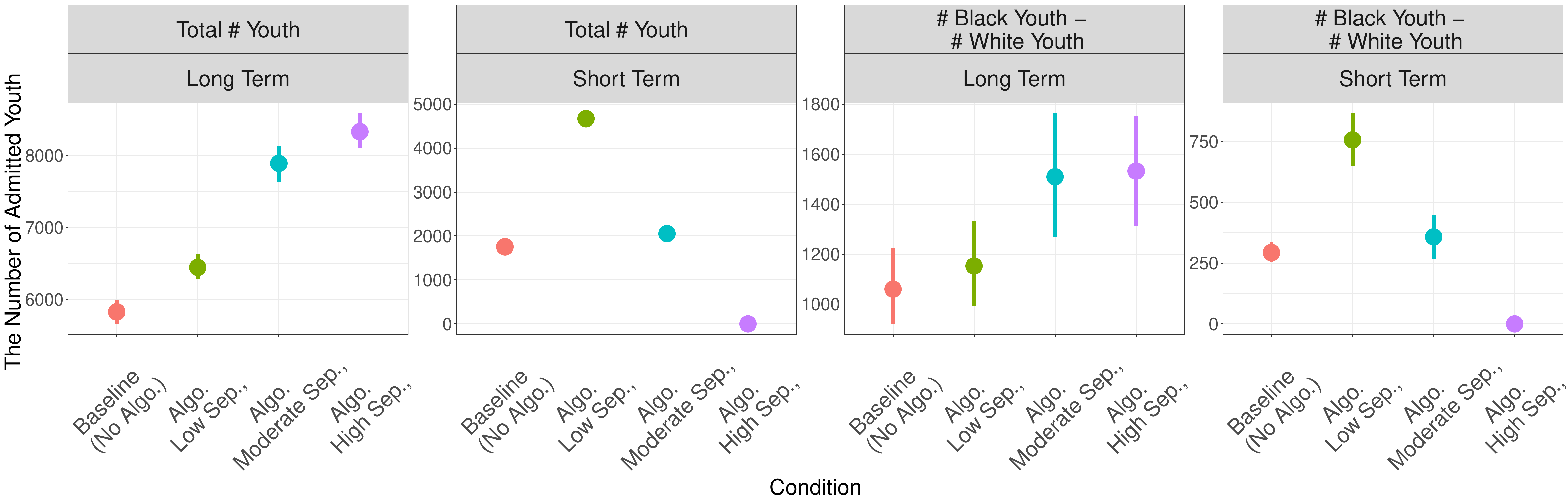}
\caption{The number of admitted long-term and short-term youth during simulated months depending on the difficulty of separability. The first two plots shows the total number of long- and short-term stays while the last two plots shows the disparities between the number of Black and white youth for long- and short-term stays.}
\label{fig:difficultyonLTST}
\end{subfigure}

\begin{subfigure}[b]{\textwidth}
    \centering
    \includegraphics[width=.5\linewidth]{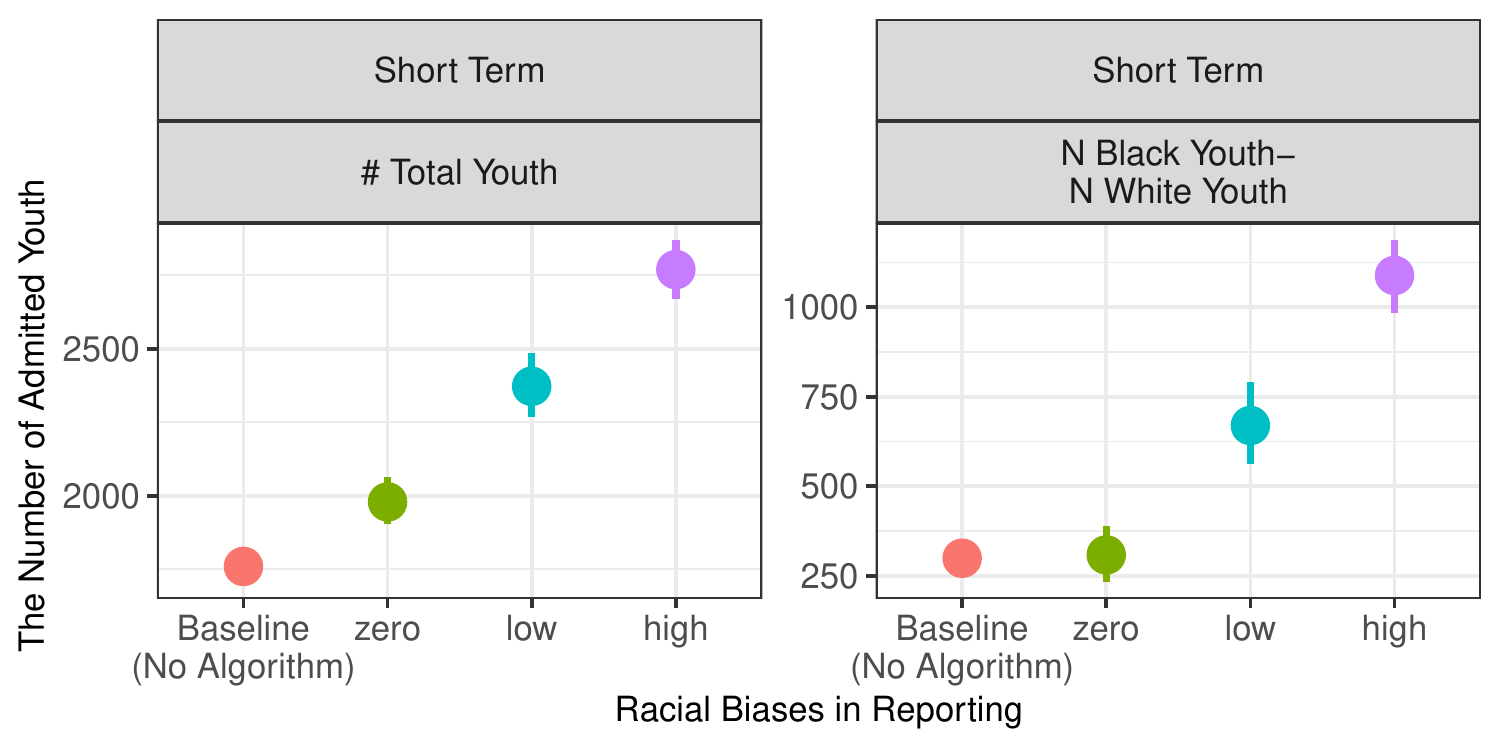}
    \caption{The number of admitted short-term youth during simulated months for different levels of racial biases in reporting in the original system. The first plot shows the total number of short term youth while the second plot shows the disparity of short term Black and white youth.}
    \label{fig:reportbias}
\end{subfigure}
\label{fig:res_two_plots}
\caption{Impact of the algorithm under varying assumptions}
\end{figure}

Overall, we find that relying on an algorithm trained as suggested by \citet{pmlr-v81-chouldechova18a} to make screen-in decision would increase the number of youth in foster care, contradicting the goals of the FFPSA. However, the impact of the algorithm varies depending on the assumed separability between profiles of youth with short-term vs. long-term stays. Figure~\ref{fig:difficulty} compares monthly forecasts of the number of Black and white youth in foster care with and without using the algorithm for making screen-in decisions. Moreover, we distinguish between different algorithm-based screen-in decisions by the ease of separability between youth with short- and long-term stays.  We see that in all conditions of separability we examined, the algorithm would increase the number of youth in foster care. However, perhaps surprisingly, the increase is significantly more pronounced when it is easy to separate long- from short- term youth. That is, the greater the assumed differences between youth with long and short-term stays, the more youth the algorithm puts into care.
 
Figure~\ref{fig:difficultyonLTST} shows a) that aggregating over youth with long- and short-term stays provides a distorted picture of the algorithm's benefits, and b) that, depending on the separability assumptions, the algorithm has non-obvious side-effects on racial disparities. More specifically, as the separability of youth with long- and short-term stays increases, the number of youth with long-term stays in foster care increases (left most plot), and so does the racial inequity for these youth (right middle plot). However, both the number of youth with short-term stays (left middle plot) \emph{and} the racial inequity (right most plot) in the number of youth with short-term stays increase as separability decreases.  Thus, under the assumptions of our model, when training an algorithm under a situation where youth who end up in care long-term are similar (as far as the algorithm is concerned) to youth who stay for a short period of time, the \emph{more} short term youth enter foster care overall. This is because the screen-in algorithm, if trained using substantiation as a proxy variable, will get better at identifying youth who the current system sees as needing long-term care, and these youth will, in turn, remain in care for longer.  Even under a seemingly better situation where there is high separability between profiles of long-term and short-term youth, algorithms will place more long-term youth into the system, contradicting the goal of FFPSA.

According to our simulation, higher levels of bias when reporting youth to CPS result in more youth in care and larger racial disparities (see Figure~\ref{fig:reportbias}).  The explanation for this phenomenon comes from the interplay of the different parts of the system: if short-term youth are more often Black, then while screen-in rates of Black youth will decrease overall, the algorithm will produce more false positives for Black youth. This effect is problematic since the reporting of youth is (a) biased \cite{harris2008decision} and (b) exogenous to the child welfare system, so a solution to this issue is needed and difficult to implement.

 
 \section{Discussion and Conclusion} \label{sec:dis}

Two major stated goals of the FFPSA are to 1) reduce unnecessary admission of youth into the foster care system and 2) provide more family-based service for youth who are in the foster care system. To achieve such goals, practitioners have been incentivized to leverage additional data collection and predictive modeling to assist decision making \cite{abdurahman2021calculating}.

Here, we construct a simulation model that considers how a particular algorithm can help to address the goals of the FFPSA, while maintaining the additional goal of racial equality. Before reviewing the main resultant claims, we find it pertinent to emphasize their limits. First, as noted above, the intervention we consider does not account for the fact that while this may not be true of tomorrow's algorithms \cite{reddenDatafiedChildWelfare2020a}, today's algorithms are often used, at best, as suggestions for screening in, rather than all-encompassing decision-makers \cite{pmlr-v81-chouldechova18a}.
Second, and related, our model greatly simplifies the ways in which algorithmic decisions are made about specific youth in foster care. Additional complexity, e.g. by creating more realistic youth profiles using advanced machine learning models, may lead to distinct conclusions. Third, certain parameterizations may be reasonably disagreed with. In particular, it is possible that higher levels of screen-in rates may lead directly to lower rates of substantiation, which violates Assumption~\ref{asu:equal}.\footnote{We thank our reviewers for pointing out this particular issue} As such, we expect, and hope, that our model is viewed primarily as a starting point for informed debate.

To this end, and with these limitations in mind, we use our model here to bring forth the point that implementing a machine learning algorithm to assist in screen-in decisions will tend instead to increase the number of youth in foster care, and to increase racial disparities in the number of Black versus white youth as well. Importantly, this increase of youth in care often comes in the form of better identification of youth that may be in genuine need of long-term foster care \textit{as defined by the current system}. In turn, while increasing the total number of youth in care, the algorithm actually decreases the number of and racial disparities in short-term stays in foster care (i.e. youth who may be candidates for deferral from foster care through prevention services). As we interpret the work of \citet{pmlr-v81-chouldechova18a}, the algorithm in this sense is accomplishing its goal---reducing racial inequality and mis-identified screen in cases \emph{as determined by the current system}.  

Our findings thus expose a contradiction between the two justifiable goals of FFPSA: 1) placing fewer youth in foster care and 2) more accurate identification of families who need more intensive services according to current measures. Namely, improving our ability to differentiate youth with long-term versus short-term stays in the current system increases the number of overall youth in foster care, but does so through the inclusion of youth who the system would likely deem to need care. Therefore, we argue that implementing algorithms that use past historical data under the new directive of the FFPSA will therefore require either 1) that we embed different notions of who needs foster care into our models or 2) that we use variables other than substantiation rates to train them.   

Equally as important, our simulation (and other analyses of the child welfare system \cite{abdurahman2021calculating}) make clear that efforts to develop algorithms for screening in youth must provide space for discussion about who, if anyone, \emph{should} be placed into foster care. Proposed algorithms shouldn't be necessarily fitted into the existing problematic loop of the social system but rather provide a new perspective to policy makers that might facilitate more effective decision-making about policy and practice at the macro level. 

 
Our work also shows that these decisions must occur with a deeper consideration of assumptions we make about the myriad and sometimes unknowable parameters of the existing system. At a fundamental level, our work shows that the assumption of substantiation decisions as a proxy for screen-in decisions made by \citet{pmlr-v81-chouldechova18a} can be problematic. Less obvious, however, is that variations in assumptions about parts of the system external to this proxy can still have important effects on model outcomes. This variation in our simulation model not simply hypothetical. Wildly different policy environments across, and even within, states mean that assumptions in one setting are quite possibly incorrect for others. These complex and hierarchical sets of assumptions benefit from the ability to systematically investigate how one change may impact others in non-obvious ways. Our simulation tool, we therefore hope, can serve as a kind of test-bed as more data becomes available to validate or inform certain assumptions in different jurisdictions, including New York state. We also hope it will assist practitioners and policy-makers in understanding the trade-offs implicit in different assumptions they are making about decisions made in their own local arena.

\begin{acks} 
This research was supported by an NSF award IIS1939579, with partial support from Amazon. The NSF and Amazon had no role in the design and conduct of the study; access and collection of data; analysis and interpretation of data; preparation, review, or approval of the manuscript; or the decision to submit the manuscript for publication. The authors declare no other financial interests.  The authors would, however, like to thank members of the UB FAI team for thoughts on this article, as well as FAccT and EAAMO reviewers who evaluated previous versions of the paper.
\end{acks}

\bibliographystyle{ACM-Reference-Format}
\bibliography{thebib}

\end{document}